\documentclass[a4paper]{spie}

\usepackage{graphicx}
\usepackage{dcolumn}
\usepackage{bm}

\begin{document}


\title{Towards redistribution laser cooling of molecular gases: Production of candidate molecules SrH by laser ablation.}



\author{P. Simon, P. Moroshkin, L. Weller, A Sa{\ss}, and M. Weitz
\skiplinehalf Institute for Applied Physics, University of Bonn,
Wegelerstr. 8, 53115 Bonn, Germany}

\pagestyle{plain}


\maketitle

\begin{abstract}
Laser cooling by collisional redistribution of radiation has been
successfully applied in the past for cooling dense atomic gases.
Here we report on progress of work aiming at the demonstration of
redistribution laser cooling in a molecular gas. The candidate
molecule strontium monohydride is produced by laser ablation of
strontium dihydride in a pressurized noble gas atmosphere. The
composition of the ablation plasma plume is analyzed by measuring
its emission spectrum. The dynamics of SrH molecular density
following the ablation laser pulse is studied as a function of the
buffer gas pressure and the laser intensity.
\end{abstract}



\section{Introduction\label{seq:Introduction}}

Collisional redistribution laser cooling is a novel technique
demonstrated for the first time by Vogl and Weitz in 2009
\cite{VoglN2009}. It is capable to laser cool "macroscopic" gas
samples of alkali/noble buffer gas mixtures at typical buffer gas
pressures of order of 100 bar \cite{VoglJMO2011,SassAPB2011}. The
cooling cycle proceeds in three steps: first, the atomic transition
of the cooled species is shifted towards lower energies by a
collision with a buffer-gas atom. At this moment, the atom absorbs a
photon from a laser beam that is red-detuned from the undisturbed
atomic transition. The two colliding atoms then spatially separate
and the transition frequency returns to its undisturbed value. In
the third step, the excited atom emits spontaneous radiation
(fluorescence) at the undisturbed frequency and returns to the
ground state. The energy difference between the absorbed and emitted
photons is taken from the kinetic energy of the colliding atoms and
the gas is thus cooled.

In the absence of inelastic collisions and saturation effects, the
cooling power density is given by
\begin{equation}
P_{cool}(r,z) =
I_{Las}(r,z)\alpha(\omega_{Las})\frac{\omega_{Fluor}-\omega_{Las}}{\omega_{Las}}
 \label{eq:CoolingPower}
\end{equation}
Here, $I_{Las}(r,z)$ is the laser intensity, $\omega_{Las}$ - the
laser frequency, $\alpha(\omega)$ - the absorption coefficient of
the medium, and $\omega_{Fluor}$ - the mean frequency of the emitted
spontaneous radiation, determined as a center of gravity of the
collision-broadened fluorescence spectrum.

The typical energy that can be removed from the dense ensemble in
one cooling cycle is of order of $k_{B}T$. Efficient cooling is thus
possible at high initial temperature $T_{0}$ and requires a large
rate of atomic collisions that can be achieved at a high buffer gas
pressure $p$. It is also important that the red-detuned laser
radiation is strongly absorbed by the gas, which requires a strong
collision broadening of the atomic spectral line and a high number
density of the absorbing atoms. The cooling effect was demonstrated
\cite{VoglN2009,VoglJMO2011,SassAPB2011,VoglSPIE2012} in
alkali-metal (Rb, K) vapors mixed with different noble gases (He,
Ar) at the initial temperature of 500-700 K and the buffer gas
pressure of 100-200 bar. Under these conditions, the sample is
optically dense even at the used detunings of 10-20 nm with respect
to the wavelength of the resonant transition of the alkali atom. The
estimated cooling power reaches 100 mW corresponding to a cooling
efficiency of $\approx$3\%, being many orders of magnitude above the
cooling efficiency in the Doppler cooling of dilute atomic gases.
The cooling mechanism is rather general and can be applied to
different species. In particular, we aim at a demonstration of the
redistribution laser cooling effect in molecular/buffer gas
mixtures.

As with other laser-cooling techniques \cite{ShumanN2010}, the main
difficulty in cooling molecules is due to the vibrational structure
of molecular transitions that makes it difficult to obtain a closed
transition for the excitation and reemission. For redistribution
laser cooling, frequent collisions with buffer gas atoms are
expected to cause thermal equilibrium of the internal rotational and
vibrational structure with the external degrees of freedom. Along
with the large efficiency of laser redistribution cooling, with an
energy removed per cooling cycle of order of $k_{B}T$, corresponding
to 1/40 eV at room temperature, the restriction to closed
transitions is expected to be loosened. For Doppler laser cooling it
was proposed \cite{DiRosaEPJD2004} to use diatomic molecules with a
diagonal Franck-Condon array. In this case, the molecule excited to
the vibronic state with the vibrational quantum number $v$, can
decay only to the lower electronic state with the same vibration
number and one obtains a closed transition suitable for the laser
cooling. Several compounds are known to possess this property,
mostly monohydrides and monohalydes of alkali-earth elements
\cite{DiRosaEPJD2004}. Among them, $A ^{2}\Pi \leftrightarrow X
^{2}\Sigma$ band of SrH has a wavelength that can be reached by a
high-power tunable Ti:sapphire laser. For this initial work we have
chosen to start with such a molecule with an almost closed
transition, aiming at a first demonstration of laser redistribution
cooling with molecules.

Unfortunately, SrH, as well as other molecules of this group, is a
chemically very active free radical and has to be produced directly
in the cooling setup, in a dense buffer-gas atmosphere. It has been
recently demonstrated
\cite{WeinsteinN1998,deCarvalhoEPJD1999,LuPCCP2011} that a very
similar compound, CaH can be produced by laser ablation of a
CaH$_{2}$ target in a cold He gas. Similar results were reported
\cite{BarryPCCP2011} also on laser ablation of SrF$_{2}$ in He that
yields SrF - another candidate molecule for redistribution laser
cooling. We therefore have chosen laser ablation as the production
method. However, obtaining a sufficient amount of SrH under high
pressure of the buffer gas has proven difficult. It should be noted
that in the existing studies laser ablation was successful at
cryogenic temperatures and relatively low buffer-gas densities.
Laser ablation of materials of this type under high buffer-gas
pressure remains largely unexplored. The present contribution
describes our approach and the current experimental status.

\section{Experiment\label{seq:Experiment}}

\subsection{Experimental setup \label{subseq:Setup}}

\begin{figure}
\begin{center}
\begin{tabular}{c}
\includegraphics[height=5.5cm]{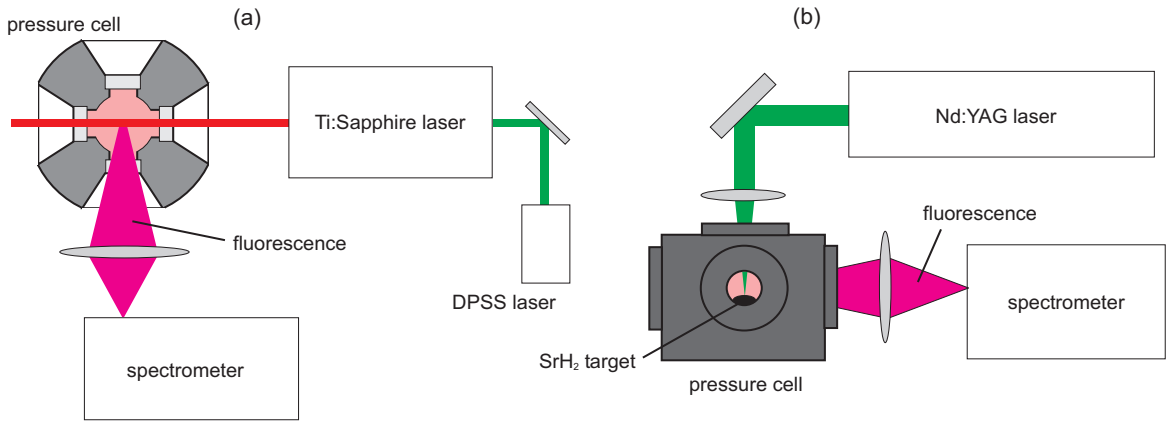}
\end{tabular}
\end{center}
\caption[example]
   { \label{fig:Setup}
Experimental setup: (a) top view, (b) side view.}
\end{figure}

Our experimental setup is sketched in Fig. \ref{fig:Setup}. We
ablate SrH$_{2}$ by a frequency-doubled pulsed Nd:YAG laser
($\lambda$ = 532 nm) with a pulse energy in the range of 2-10 mJ and
a repetition rate of 10 Hz. The ablation target is prepared by
pressing the SrH$_{2}$ powder by a hydraulic press developing a
pressure of $\approx$400 MPa. The target is installed at the bottom
of a high-pressure sample cell filled with He or Ar buffer gas at a
pressure of 0.5 - 10 bar. The cell is made of stainless steel with
five sapphire windows. Four horizontal windows are used for the
diagnostics and the top window for the ablation. The Nd:YAG laser
beam is focused by a lens with a focal distance of 12 cm, mounted
above the cell on a translation stage. We have obtained the most
stable ablation yield with the lens positioned $\approx$ 10 cm above
the target surface. We estimate the beam cross section at the target
surface as $S \approx$ 1 mm$^{2}$. With more tight focusing, we
observe that the laser beam creates a crater in the target surface
on the time scale of several minutes. The ablation yield
subsequently becomes strongly reduced.

In order to analyze the composition of the created ablation plasma,
we collect the fluorescence emitted by the plume through one of the
horizontal windows. The light is then coupled into a fiber and
brought to a grating spectrometer with a resolution of 1 nm and
covering a spectral range of 300-1000 nm.

We detect SrH molecules via their laser-induced fluorescence excited
by a radiation of a cw Ti:sapphire laser pumped with a
frequency-doubled diode-pumped solid state (DPSS) laser. The
Ti:sapphire laser is tuned into resonance with the $v=0 \rightarrow
v =0$ vibrational component of the $X$ $^{2}\Sigma_{1/2} \rightarrow
A$ $^{2}\Pi_{3/2}$ band at 733 nm. The laser beam passes through a
pair of horizontal windows, 1-2 mm above the target surface. The
fluorescence is collected at right angle with respect to the laser
beam. The fluorescence is separated from the scattered laser beam
light and ablation plume emission by means of a home-built
monochromator utilizing a diffraction grating of 1200 grooves/mm and
having a resolution of about 2 nm. The fluorescence signal is
detected by a photomultiplier tube (PMT) and recorded by a digital
oscilloscope. We detect the fluorescence at the $v'=0 \rightarrow
v=0$ vibrational component of the $A$ $^{2}\Pi_{1/2} \rightarrow X$
$^{2}\Sigma_{1/2}$ band at 745 nm. The $A$ $^{2}\Pi_{1/2}(v'=0)$
state lies 300 cm$^{-1}$ below the laser-excited $A$
$^{2}\Pi_{3/2}(v'=0)$ state and is populated by atomic collisions.

\subsection{Experimental results \label{subseq:Results}}

\begin{figure}
\begin{center}
\begin{tabular}{c}
\includegraphics[height=8cm]{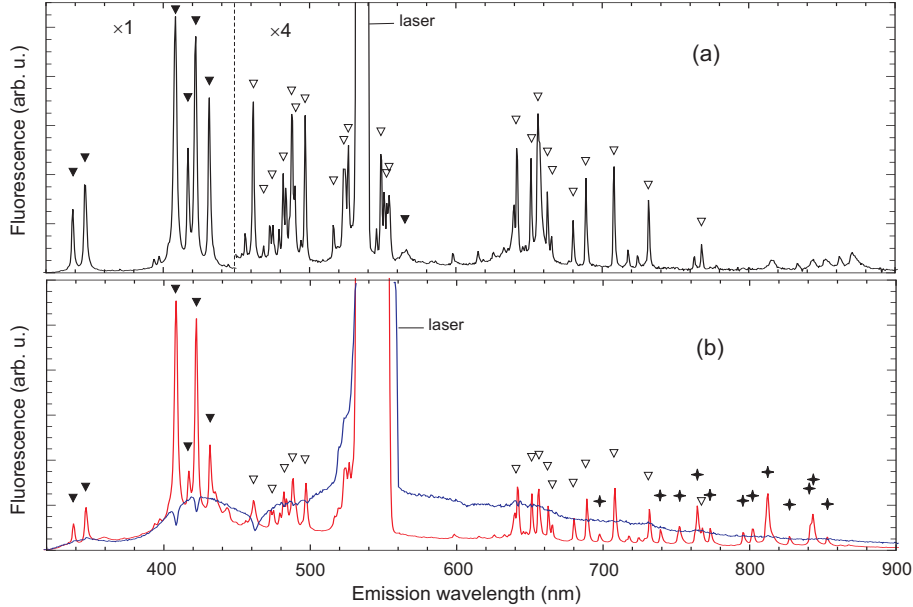}
\end{tabular}
\end{center}
\caption[example]
   { \label{fig:AblationSpectra}
Emission spectra of the ablation plasma plume in vacuum (a) and in
the argon atmosphere (b). In (b), red line - $p$ = 2 bar, blue line
- $p$ = 100 bar. Filled and empty triangles mark the transitions of
Sr$^{+}$ ions and Sr atoms, respectively, crosses mark the Ar
lines.}
\end{figure}

Typical spectra of the emission produced by the ablation plume at
different buffer gas (argon) pressures are shown in Fig.
\ref{fig:AblationSpectra}. The wavelengths of the observed lines and
their assignments to the corresponding electronic transitions of Sr
atoms and Sr$^{+}$ ions are collected in Table \ref{tab:SrI} and
Table \ref{tab:SrII}, respectively. The near infrared part of the
spectrum is dominated by Ar lines. In the violet part of spectrum we
observe six strong lines of Sr$^{+}$ ion. The spectral range of 450
- 700 nm contains mostly lines of the Sr atom. The $H_{\alpha}$ line
is visible at 656.3 nm. All other lines of atomic hydrogen overlap
with much stronger lines of Sr and Sr$^{+}$ and could not be
observed. The atomic and ionic lines are strongly broadened by
collisions. Increasing the Ar pressure leads to a stronger
broadening and to the appearance of a continuous emission spectrum
that extends over a wavelength range from 400 to 900 nm.

At high buffer gas pressure, the ablation plume becomes more compact
since the ablation products experience more collisions with buffer
gas atoms and thermalize at shorter distance from the surface. As a
result, we obtain a higher density of the ablation products. It
leads to a reabsorption of the emitted radiation by the atoms and
ions in the plume. Those lines that terminate at the ground state
therefore appear as dips in the emission spectrum. In Fig.
\ref{fig:AblationSpectra} (b), one can see pronounced
self-absorption dips at the wavelengths of the $5^{2}P_{3/2} -
5^{2}S_{1/2}$ and $5^{2}P_{1/2} - 5^{2}S_{1/2}$ lines of Sr$^{+}$ at
407.8 and 421.5 nm, respectively and of the $5^{1}P_{1} -
5^{1}S_{0}$ transition of Sr at 460.7 nm.

\begin{table}
\caption{Electronic transitions of Sr atoms observed in the
fluorescence emitted by the laser-ablation plume.} \label{tab:SrI}
\begin{center}
\begin{tabular}{c|c|c||c|c|c}
 & transition & $\lambda$ (nm) &  & transition & $\lambda$ (nm) \\\hline\hline
 5$s6f$ - $5s4d$ & $^{1}F_{3}$ - $^{1}D_{2}$ & 440.6 & & $^{3}D_{1}$ - $^{3}P_{0}$ & 483.2 \\
 5$s7s$ - $5s5p$ & $^{3}S_{1}$ - $^{3}P_{2}$ & 443.8 & 5$s5d$ - $5s5p$ & $^{3}D_{2}$ - $^{3}P_{1}$ & 487.2 \\
 5$s5p$ - $5s^{2}$ & $^{1}P_{1}$ - $^{1}S_{0}$ & 460.7 & & $^{3}D_{3}$ - $^{3}P_{2}$ & 496.2 \\
 5$s5f$ - $5s4d$ & $^{1}F_{3}$ - $^{1}D_{2}$ & 467.8 & & $^{1}D_{2}$ - $^{1}P_{1}$ & 767.3 \\\hline
  & $^{3}P_{2}$ - $^{3}P_{1}$ & 472.2 & & $^{1}P_{1}$ - $^{1}D_{2}$ & 475.5 \\
  & $^{3}P_{1}$ - $^{3}P_{0}$ & 474.2 & & $^{3}P_{1}$ - $^{3}D_{1}$ & 522.2 \\
 5$p^{2}$ - $5s5p$ & $^{3}P_{1}$ - $^{3}P_{1}$ & 478.4 & & $^{3}P_{1}$ - $^{3}D_{2}$ & 523.8 \\
  & $^{3}P_{2}$ - $^{3}P_{2}$& 481.2 & & $^{3}P_{2}$ - $^{3}D_{3}$ & 525.7 \\
  & $^{1}D_{2}$ - $^{1}P_{1}$& 655.0 & & $^{3}D_{3}$ - $^{3}D_{2}$ & 545.1 \\\hline
  & $^{3}F_{2}$ - $^{3}D_{1}$ & 485.5 & & $^{3}D_{3}$ - $^{3}D_{3}$ & 548.1 \\
 5$s4f$ - $5s4d$ & $^{3}F_{4}$ - $^{3}D_{3}$ & 489.2 & & $^{3}D_{2}$ - $^{3}D_{2}$ & 550.4 \\
  & $^{1}F_{3}$ - $^{1}D_{2}$ & 515.6 & 4$d5p$ - $5s4d$ & $^{3}D_{1}$ - $^{3}D_{1}$ & 552.2 \\\hline
  & $^{3}P_{2}$ - $^{3}D_{2}$ & 634.6 & & $^{3}D_{1}$ - $^{3}D_{2}$ & 554.0 \\
 5$s6p$ - $5s4d$ & $^{3}P_{1}$ - $^{3}D_{2}$ & 638.8 & & $^{3}F_{4}$ - $^{3}D_{3}$ & 640.8\\
  & $^{1}P_{1}$ - $^{1}D_{2}$ & 716.7 & & $^{3}F_{3}$ - $^{3}D_{2}$ & 650.4 \\
  & $^{3}P_{2}$ - $^{1}D_{2}$ & 723.2 & & $^{3}F_{2}$ - $^{3}D_{1}$ & 661.7 \\\hline
  & $^{3}S_{1}$ - $^{3}P_{0}$ & 679.1 & & $^{3}F_{2}$ - $^{3}D_{2}$ & 664.3 \\
 5$s6s$ - $5s5p$ & $^{3}S_{1}$ - $^{3}P_{1}$ & 687.8 & & $^{1}D_{2}$ - $^{1}D_{2}$ & 730.9 \\
  & $^{3}S_{1}$ - $^{3}P_{2}$ & 707.0 & & $^{3}F_{2}$ - $^{1}D_{2}$ & 762.1 \\\hline
\end{tabular}
\end{center}
\end{table}

\begin{table}
\caption{Electronic transitions of Sr$^{+}$ ions observed in the
fluorescence emitted by the laser-ablation plume.} \label{tab:SrII}
\begin{center}
\begin{tabular}{c|c|c}
 & transition & $\lambda$ (nm) \\\hline\hline
4$p^{6}5d$ - $4p^{6}5p$ & $^{2}D_{3/2}$ - $^{2}P_{1/2}$ & 338.1 \\
 & $^{2}D_{5/2}$ - $^{2}P_{3/2}$ & 346.4 \\\hline
4$p^{6}5p$ - $4p^{6}5s$ & $^{2}P_{3/2}$ - $^{2}S_{1/2}$ & 407.8\\
 & $^{2}P_{1/2}$ - $^{2}S_{1/2}$ & 421.5\\\hline
4$p^{6}6s$ - $4p^{6}5s$ & $^{2}S_{1/2}$ - $^{2}P_{1/2}$ &
416.2 \\
 & $^{2}S_{1/2}$ - $^{2}P_{3/2}$ &
430.5 \\\hline
\end{tabular}
\end{center}
\end{table}

The observed fluorescence spectra demonstrate high densities of Sr
atoms and Sr$^{+}$ ions produced by laser decomposition of
SrH$_{2}$. However, no spectral features of SrH or SrH$_{2}$
molecules could be detected in the plasma plume spectra. These
observations suggest that the laser ablation leads to a complete
decomposition of SrH$_{2}$ molecules and to the strong ionization of
the dissociation products. In order to search for the SrH molecules
in the ablation plume at larger delay-times with respect to the
laser pulse, we have excited the gas with a cw Ti:sapphire laser
tuned in resonance with the $X$ $^{2}\Sigma_{1/2}(v=0) \rightarrow
A$ $^{2}\Pi_{3/2}(v'=0)$ band at 733 nm. To avoid a strong
background from scattered incident laser radiation, we monitor the
fluorescence emitted by the $A$ $^{2}\Pi_{1/2}(v'=0)$ state that is
located energetically below the $A$ $^{2}\Pi_{3/2}(v'=0)$ state and
is populated by collisions with buffer gas atoms.

\begin{figure}
\begin{center}
\begin{tabular}{c}
\includegraphics[height=5cm]{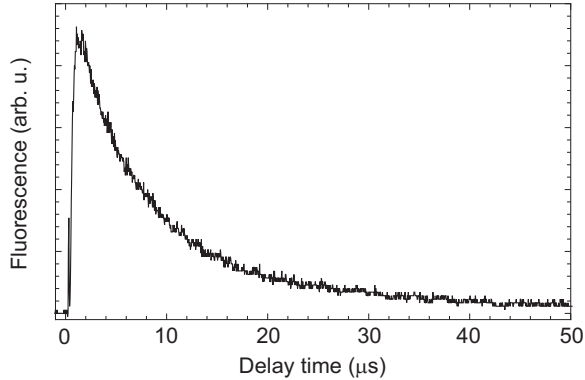}
\end{tabular}
\end{center}
\caption[example]
   { \label{fig:TimeResolved}
Time-resolved laser-induced fluorescence signal of SrH following the
ablation-laser pulse. He buffer gas, $p$ = 2 bar, excitation by cw
laser radiation at $\lambda_{exc}$ = 733.6 nm, fluorescence at 745
nm.}
\end{figure}

No SrH fluorescence could so far be detected in Ar buffer gas.
However, we have observed a strong laser-induced fluorescence signal
following the ablation of SrH$_{2}$ target in He atmosphere. A
typical time-resolved fluorescence signal is shown in Fig.
\ref{fig:TimeResolved}. The broadband emission of the ablation
plasma plume is produced within a time interval of 200 ns
immediately after the ablation laser pulse. It has been measured
separately and then subtracted from the fluorescence signal. The
radiative life time of the $A$ $^{2}\Pi_{1/2}$ state of SrH, as
measured by Berg \textit{et al.} \cite{BergCPL1996}, is 34 ns and
that of $A$ $^{2}\Pi_{3/2}$ is even shorter \cite{BergCPL1996}. The
observed rise and decay of the SrH fluorescence excited by the
resonant cw laser thus reflect the dynamics of the number density of
the ground state SrH molecules. The signal rises after the ablation
pulse, reaches a maximum at a delay time of $\approx$1 $\mu$s and
then decays exponentially with a decay-time of 8.6 $\mu$s.

\begin{figure}
\begin{center}
\begin{tabular}{c}
\includegraphics[height=5.5cm]{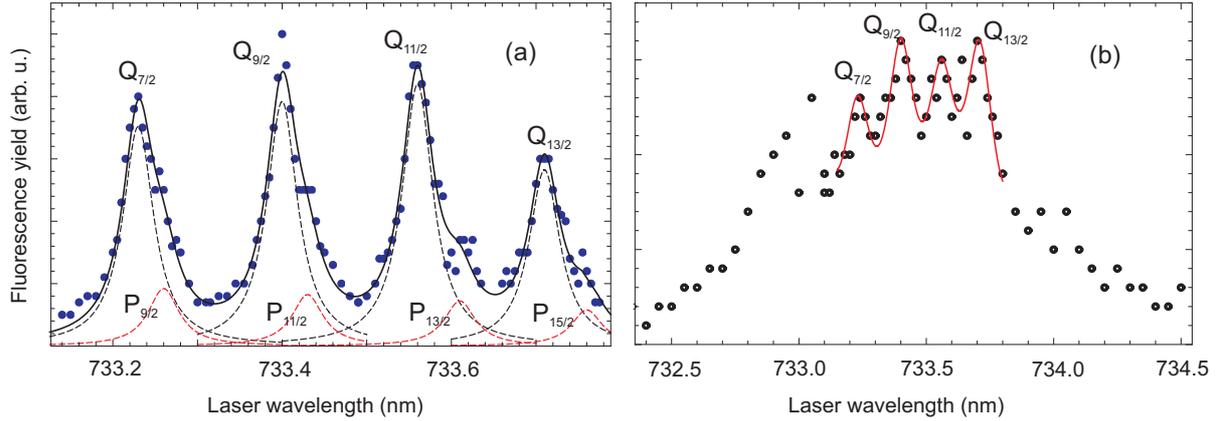}
\end{tabular}
\end{center}
\caption[example]
   { \label{fig:ExcSpec}
Excitation spectrum of the $X$ $^{2}\Sigma_{1/2}(v=0) \rightarrow A$
$^{2}\Pi_{3/2}(v'=0)$ band of SrH in He buffer gas. Dots -
experimental data, solid line - fit to the data. (a) $p$ = 2 bar;
(b) $p$ = 10 bar. In (a), dashed lines show individual contributions
of different transitions from P and Q branches, labeled with the $J$
quantum numbers of the corresponding lower states.}
\end{figure}

The absorption spectrum of the $X$ $^{2}\Sigma_{1/2}(v=0)
\rightarrow A$ $^{2}\Pi_{3/2}(v'=0)$ band of SrH was measured by
scanning the excitation laser in the vicinity of $\lambda_{exc}$ =
733 nm and measuring the fluorescence yield at $\lambda_{fluor}$ =
748 nm, as a function of $\lambda_{exc}$. At low buffer gas
pressures the rotational structure could be resolved, as shown in
Fig. \ref{fig:ExcSpec}(a). One can see four resolved spectral
components of the $Q$-branch, each of them overlapping with a much
weaker line from the $P$-branch. The experimental spectrum is fitted
with a sum of 8 Lorentzian curves having the same spectral width and
centered at the positions of the rovibronic lines accordingly to the
data of \cite{WatsonPR1932,AppelbladPS1986}. The individual
contributions of 8 lines are also shown by dashed lines. The lines
are labeled by the total angular momentum quantum numbers $J$ of the
corresponding lower states. This fit allows us to determine the
spectral widths of the molecular lines broadened by collisions with
the buffer gas atoms. At $p$ = 2 bar, the FWHM spectral width of
each line is 24.7$\pm$0.7 GHz, which is much larger than the Doppler
width at room temperature.

A typical excitation spectrum at a buffer gas pressure of 10 bar is
shown in Fig. \ref{fig:ExcSpec}(b). At high pressure, the line
broadening increases and the rotational structure can not be
resolved any more. The whole band has a FWHM spectral width of
$\approx$1 nm and has broad spectral wings.

\section{Discussion\label{seq:Discussion}}

Our measurements demonstrate the feasibility of the production of
SrH molecules by laser ablation of SrH$_{2}$ in a helium buffer gas
atmosphere. The dominant species in the ablation plume are Sr atoms
and Sr$^{+}$ ions. During the laser pulse their number density
reaches very high values, as is evidenced by the reabsorption of
their fluorescence. On the other hand, SrH molecules could be
detected in relatively smaller amounts and their observed lifetime
after the ablation pulse is in the microseconds range. Surprisingly,
no SrH spectral lines could so far be observed in argon buffer gas.
A possible explanation for this would be that collisions with Ar
atoms or some other ablation products effectively quench the excited
states of SrH.

The measurements of the excitation spectrum of the $X$
$^{2}\Sigma_{1/2}(v=0) \rightarrow A$ $^{2}\Pi_{3/2}(v'=0)$ band
provide information about the collision line broadening that is very
important for the planned molecular redistribution cooling
experiment. At buffer gas pressures above 10 bar, the
collision-broadened rovibronic lines overlap strongly and form a
continuous absorption spectrum. The $X$ $^{2}\Sigma_{1/2}(v=0)
\rightarrow A$ $^{2}\Pi_{3/2}(v'=0)$ band itself is not suitable for
the redistribution cooling experiment because the laser-excited
molecules in the $A$ $^{2}\Pi_{3/2}(v'=0)$ state undergo
radiationless transitions towards the lower lying $A$
$^{2}\Pi_{1/2}(v'=0)$ state. The energy difference of $\approx$300
cm$^{-1}$ is transferred into the kinetic energy and will lead to
heating of the gas. On the other hand, redistribution laser cooling
should be possible at the $X$ $^{2}\Sigma_{1/2}(v=0) \rightarrow A$
$^{2}\Pi_{1/2}(v'=0)$ band that is centered around 748 nm, which is
expected to have a similar absorption spectrum.

The yield of SrH molecules achieved in the present experiment is
relatively low compared to typical atomic densities in the so far
carried out redistribution cooling experiments
\cite{VoglN2009,VoglJMO2011,SassAPB2011}. In fact, we could so far
not observe a measurable attenuation of the resonant laser beam
crossing the sample cell. WE expect that the yield can be
substantially increased by increasing the laser power with a
corresponding increase of the laser beam diameter at the target
surface in order to keep the laser intensity fixed. We have obtained
the best results with the laser fluences of order of 0.5 J/cm$^{2}$.

Another important observation is that of a decay of the molecular
number density on the time-scale of 10 $\mu$s. Although the
diffusion coefficient for SrH in He is not known, it is highly
unlikely that a molecule can leave the observation volume and reach
the cell wall in such a short time. Other loss mechanisms may
include a chemical reaction in collisions with Sr or H atoms, the
formation of clusters or condensation. An increase of the He
pressure is expected to slow down the diffusion of SrH molecules. On
the other hand, it will also make thermalization of the ablated
species faster and confine the ablation plume close to the target
surface. This in turn may lead to condensation of the ablated
molecules into clusters or on the target surface. A similar behavior
of laser-ablated Au and Li atoms in He buffer gas was reported by
Sushkov and Budker \cite{SushkovPRA2008}. At low He densities they
observed a diffusion-controlled dynamics of the atomic density with
a typical decay time of order of several milliseconds. However, at
densities above 10$^{18}$ cm$^{-3}$, an additional loss mechanism
starts to dominate that leads to a decrease of the decay time and of
the ablation yield with increasing buffer gas density. The present
work explored the range of He densities of 10$^{19}$ - 10$^{21}$
cm$^{-3}$. The observed yield and the short decay time are in
qualitative agreement with the results of \cite{SushkovPRA2008}.

\section{Summary\label{seq:Summary}}

To conclude, we report on a study of the production of SrH molecules
in noble buffer gas conditions at high pressures for redistribution
laser cooling. The molecular sample was produced by laser ablation
in a buffer gas cell. We investigate the spectral broadening of the
$X$ $^{2}\Sigma_{1/2}(v=0) \rightarrow A$ $^{2}\Pi_{3/2}(v'=0)$ band
of SrH by atomic collisions and the dynamics of the molecular
density following the ablation laser pulse. For the future, we plan
to carry out laser redistribution cooling of SrH molecules under
high pressure noble buffer gas conditions.


\begin{acknowledgments}
This work was supported by the Deutsche Forschungsgemeinschaft,
grant No: We 1748-15.
\end{acknowledgments}

\bibliography{AblationSrH}

\begin{thebibliography}{10}

\bibitem{VoglN2009}
Vogl, U. and Weitz, M., ``Laser cooling by collisional redistribution of
  radiation,'' {\em Nature}~{\bf 461},  70--73 (2009).

\bibitem{VoglJMO2011}
Vogl, U., Sa{\ss}, A., Ha{\ss}elmann, S., and Weitz, M., ``Collisional
  redistribution laser cooling of a high-pressure atomic gas,'' {\em J. Mod.
  Optics}~{\bf 58}(15),  1300--1309 (2011).

\bibitem{SassAPB2011}
Sa${\ss}$, A., Vogl, U., and Weitz, M., ``Laser cooling of potassium-argon gas
  mixture using collisional redistribution of radiation,'' {\em Appl. Phys.
  B}~{\bf 102},  503--507 (2011).

\bibitem{VoglSPIE2012}
Vogl, U., Sa{\ss}, A., and Weitz, M., ``Laser cooling of dense rubidium-noble
  gas mixtures via collisional redistribution of radiation,'' {\em Proc.
  SPIE}~{\bf 8275},  827508 (2012).

\bibitem{ShumanN2010}
Shuman, E.~S., Barry, J.~F., and DeMille, D., ``Laser cooling of a diatomic
  molecule,'' {\em Nature}~{\bf 467},  820--823 (2010).

\bibitem{DiRosaEPJD2004}
{D}i Rosa, M.~D., ``Laser cooling molecules,'' {\em Eur. Phys. J. D}~{\bf 31},
  395--402 (2004).

\bibitem{WeinsteinN1998}
Weinstein, J.~D., de~Carvalho, R., Guillet, T., Friedrich, B., and Doyle,
  J.~M., ``Magnetic trapping of calcium monohydride molecules at millikelvin
  temperatures,'' {\em Nature}~{\bf 395},  148--150 (1998).

\bibitem{deCarvalhoEPJD1999}
de~Carvalho, R., Doyle, J.~M., Friedrich, B., Guillet, T., Kim, J., Patterson,
  D., and Weinstein, J.~D., ``Buffer-gas loaded magnetic traps for atoms and
  molecules: {A} primer,'' {\em Eur. Phys. J. D}~{\bf 7},  289--309 (1999).

\bibitem{LuPCCP2011}
Lu, H.~I., Rasmussen, J., Wright, M.~J., Patterson, D., and Doyle, J.~M., ``A
  cold and slow molecular beam,'' {\em Phys. Chem. Chem. Phys.}~{\bf 13},
  18986--18990 (2011).

\bibitem{BarryPCCP2011}
Barry, J.~F., Shuman, E.~S., and DeMille, D., ``A bright, slow cryogenic
  molecular beam source for free radicals,'' {\em Phys. Chem. Chem. Phys.}~{\bf
  13},  18936--18947 (2011).

\bibitem{BergCPL1996}
Berg, L.~E., Ekvall, K., Hishikawa, A., Kelly, S., and McGuinness, C., ``Laser
  spectroscopy of {SrH}. {T}ime-resolved measurements of the {$A^{2}\Pi$}
  state,'' {\em Chem. Phys. Lett.}~{\bf 255},  419--424 (1996).

\bibitem{WatsonPR1932}
Watson, W.~W. and Fredrickson, W.~F., ``The spectrum of strontium hydride,''
  {\em Phys. Rev.}~{\bf 39},  765--777 (1932).

\bibitem{AppelbladPS1986}
Appelblad, O., Klynning, L., and Johns, J. W.~C., ``Fourier transform
  spectroscopy of {SrH}: The {$A-X$} and {$B-X$} band systems,'' {\em Phys.
  Scripta}~{\bf 33},  415--419 (1986).

\bibitem{SushkovPRA2008}
Sushkov, A.~O. and Budker, D., ``Production of long-lived atomic vapor inside
  high-density buffer gas,'' {\em Phys. Rev. A}~{\bf 77},  042707 (2008).

\end{thebibliography}
\bibliographystyle{spiebib}

\end{document}